\documentclass[prl, twocolumn]{revtex4}
\usepackage{graphicx}
\usepackage{amsmath}
\usepackage{amssymb}
\usepackage{bm}
\usepackage{color}
\usepackage{hyperref}
\usepackage{tabularx}
\usepackage{multirow}
\usepackage{braket}
\usepackage{xcolor}

\begin{document}

\title{Magnetoelectric Effects in Gyrotropic Superconductors}
\author{Wen-Yu He and K. T. Law}\thanks{Correspondence address: phlaw@ust.hk}
\affiliation{Department of Physics, Hong Kong University of Science and Technology, Clear Water Bay, Hong Kong, China}
\date{\today}
\pacs{}

\begin{abstract}
{The magnetoelectric effect (or Edelstein effect) in non-centrosymmetric superconductors states that a supercurrent can induce spin magnetization. This is an intriguing phenomenon which has potential applications in superconducting spintronic devices. However, the original Edelstein effect only applies to superconductors with polar point group symmetry. In recent years, many new noncentrosymmetric superconductors have been discovered such as superconductors with chiral lattice symmetry and superconducting transition metal dichalcogenides with various lattice structures. In this work, we provide a general framework to describe the supercurrent induced magnetization in these newly discovered superconductors with gyrotropic point groups.}
\end{abstract}

\maketitle
\section{Introduction}

\begin{figure*}
\centering
\includegraphics[width=\textwidth]{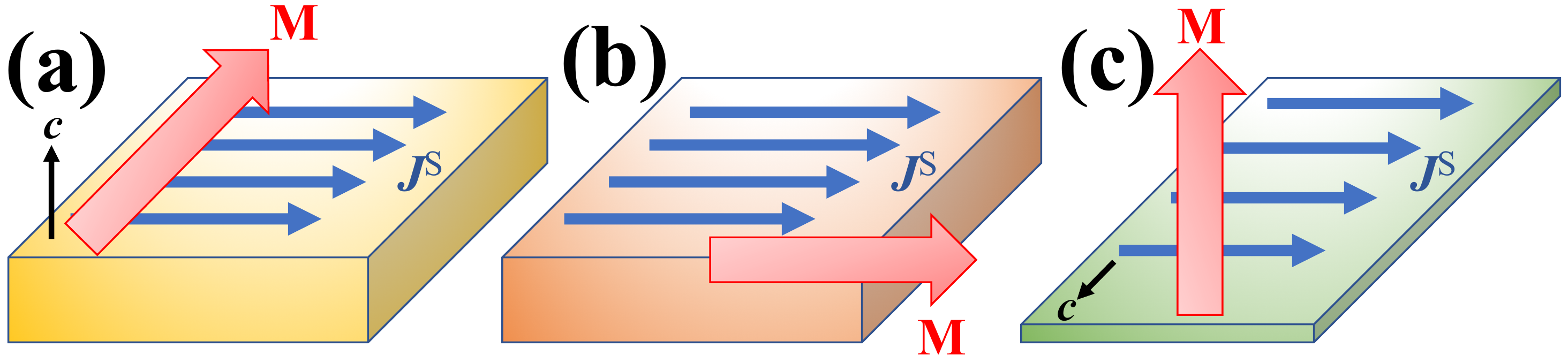}
\caption{The supercurrent driven magnetization in (a) polar superconductors with Rashba SOC, (b) superconducting chiral crystals belonging to T and O point groups, and (c) quasi-2D superconducting TMDs with polar axis lying inside the atomic plane. In polar superconductors with Rashba SOC, the supercurrent driven magnetization $\bm{M}$ is the transverse response as $\bm{M}\propto\bm{c}\times\bm{J}^{\textrm{S}}$. In superconducting chiral crystals of T and O point group, the magnetization arising from supercurrent is parallel to the supercurrent direction and the magnetoelectric effect becomes a longitudinal response. In quasi-2D superconducting TMDs with C$_1$, C$_2$, C$_{1v}$ or C$_{2v}$ point group symmetry, the polar axis $\bm{c}$ lies inside the atomic plane and allows the in-plane supercurrent to induce the out of plane magnetization.}\label{figure1}
\end{figure*}

In noncentrosymmetric metals, spin and momentum of electrons are coupled such that the deformation of the Fermi surface due to dissipative current can polarize electron spins and this is called the magnetoelectric effect or the Edelstein effect~\cite{Levitov1, Edelstein1, Fiebig}. More interestingly, in some noncentrosymmetric superconductors, supercurrents can also give rise to a static magetization without dissipation~\cite{Fiebig, Levitov2, Edelstein2, Edelstein3, Yip, Fujimoto}. And conversely, a static magnetization can drive local supercurrents~\cite{Yip, Samokhin, Agterberg, Feigelman, Sigrist1, Mironov}. As pointed out by Edelstein~\cite{Edelstein2, Edelstein3}, in superconductors with a polar axis $\bm{c}$ and Rashba spin-orbit coupling (SOC), the magnetization $\bm{M}$ induced by supercurrents can be expressed as $\bm{M}\propto\bm{c}\times\bm{J}^{\textrm{S}}$~\cite{Edelstein2} where $\bm{J}^{\textrm{S}}$ is the supercurrent density. In such Rashba superconductors, the induced magnetization is always perpendicular to the direction of supercurrent and the polar axis, as is shown in Fig. \ref{figure1} (a).

In recent years, many noncentrosymmetric superconductors such as superconductors with chiral lattice structures~\cite{Salamon, Hirata, Prozorov, Carnicom}, and superconducting transition metal dichalcogenides (TMDs)~\cite{Jianting, Saito, Mak, Cava, Sajadi, Sanfeng, Guangtong} are discovered. Since the newly emergent noncentrosymmetric superconductors have different crystal symmetries, the SOC in these materials have different forms. For example, superconductors with point group symmetry T and O, which belong to chiral (or enantiomorphic) point groups, have isotropic SOC in the form of $v\bm{p}\cdot\bm{\sigma}$, where $\bm{p}$ is momentum, $\bm{\sigma}$ is the Pauli matrices and $v$ is the SOC strength~\cite{Guoqing}. On the other hand, multilayer 1Td-structure WTe$_2$ and MoTe$_2$ possess anisotropic SOC~\cite{Guangtong}. For atomically thin monolayer 2H-MoS$_2$ and 2H-NbSe$_2$,  Ising SOC~\cite{Jianting, Saito, Mak, Noah, Benjamin, Wenyu} is present which pins electron spins to the out-of-plane directions. These different forms of SOC are expected to cause unconventional magnetoelectric effects different from the one in Fig. \ref{figure1} (a) for Rashba superconductors. However, a general understanding of the magnetoelectric effects of all these noncentrosymmetric superconductors is lacking. In this work, through linear response theory and group theory analysis, we provide a general and powerful way to understand magnetoelectric effects in noncentrosymmetric superconductors. Importantly, we point out that in superconducting chiral crystals and superconducting quasi-2D TMDs, applying supercurrent can give novel magnetization shown in Fig. \ref{figure1} (b) and (c) respectively.

First of all, we note that among the superconductors within the 21 noncentrosymmetric point groups, the magnetoelectric effect is generally non-zero only for the ones belonging to the 18 gyrotropic point groups~\cite{gyrotropy, Moore}, we call these superconductors gyrotropic superconductors.  The explicit forms of the magnetoelectric pseudotensors for the 18 gyrotropic point groups are listed in Table \ref{magnetoelectric_pseudotensor}. According to Table \ref{magnetoelectric_pseudotensor}, many of the novel magnetoelectric response of noncentrosymmetric superconductors can be identified immediately. 

For example, for materials belonging to the T and O point groups, such as Li$_2$Pd$_3$B, Li$_2$Pt$_3$B~\cite{Salamon, Hirata}, and Mo$_3$Al$_2$C~\cite{Prozorov}, the magnetoelectric effect is purely longitudinal, meaning that the induced spin magnetization is always parallel to the direction of the supercurrent, as is shown in Fig. \ref{figure1} (b). This is a quantum analogue of classical solenoids in which the induced magnetic field is parallel to the current directions. We further point out that, for quasi-two-dimensional materials, the only symmetries which allow to have induced spin magnetization perpendicular to the atomic plane are the ones which have the polar axis lying inside the atomic plane. The newly discovered multilayer 1Td-structure WTe$_2$~\cite{Sajadi, Sanfeng} and MoTe$_2$~\cite{Guangtong} naturally fulfils this condition so that there exists the in-plane supercurrent induced out of plane magnetization seen from Fig. \ref{figure1} (c).

On the other hand, Ising superconductors such as 2H-structure MoS$_2$ and NbSe$_2$, the magnetoelectric effect is indeed zero due to their D$_{3h}$ point group, even though the SOC in these materials are particularly strong. These 2H-TMDs are interesting examples of noncentrosymmetric superconductors which give rise to zero magnetoelectric response. Importantly, under uniaxial strain which reduce D$_{3h}$ to C$_{2v}$, a purely out-of-plane magnetization can be induced by a supercurrent in 2H-TMD. Therefore, our results, as summerized in Table \ref{magnetoelectric_pseudotensor}, can be used to provide guiding principles to generate and manipulate magnetoelectric responses in superconductors.

The rest of the paper is organised as follows. First, we construct the microscopic model for the magnetoelectric effect and show that the induced spin magnetization is related to the supercurrent by a rank-two pseudotensor $T_{ij}$. Second, we analyse the symmetry properties of $T_{ij}$ under different point group operations and list the general form of $T_{ij}$ in Table \ref{magnetoelectric_pseudotensor}. Third, we discuss the application of Table \ref{magnetoelectric_pseudotensor} in the understanding of several interesting superconductors with gyrotropic point group, including the superconductor with chiral lattice symmetry and superconducting TMDs. Finally, we demonstrate how an unconventional magnetoelectric response can be generated by strain in non-gyrotropic superconductors using 2H-structure TMD as an example.

\section{Results}

\subsection{Theory for the superconducting magnetoelectric effect}
To study a noncentrosymmetric superconductor which carries a supercurrent, we consider the Bogliubov de Gennes Hamiltonian
\begin{align}
H\left(\bm{p}, \bm{q}\right)=&\begin{pmatrix}
\xi_{\bm{p}+\frac{1}{2}\bm{q}}+\bm{g}_{\bm{p}+\frac{1}{2}\bm{q}}\cdot\bm{\sigma} & i\Delta_{\bm{q}}\sigma_y\\ 
-i\Delta_{\bm{q}}^\ast\sigma_y & -\xi_{-\bm{p}+\frac{1}{2}\bm{q}}-\bm{g}_{-\bm{p}+\frac{1}{2}\bm{q}}\cdot\bm{\sigma}^\ast
\end{pmatrix}.
\end{align}
Here the Nambu basis is $\left[c^\dagger_{\bm{p}+\frac{1}{2}\bm{q}, \uparrow}, c^\dagger_{\bm{p}+\frac{1}{2}\bm{q}, \downarrow}, c_{-\bm{p}+\frac{1}{2}\bm{q}, \uparrow}, c_{-\bm{p}+\frac{1}{2}\bm{q}, \downarrow}\right]$ with $c^{\left(\dagger\right)}_{\bm{p}', s}$ the annihilation (creation) operator at the momentum $\bm{p}'$ and spin index $s=\uparrow/\downarrow$.  The kinetic energy of an electron with momentum $\bm{p}$ is described by $\xi_{\bm{p}}$. The vector $\bm{g}_{\bm{p}}$ describes the SOC of the material and its specific form is determined by the lattice symmetry, and $\bm{\sigma}$ denotes the Pauli matrices. From the definition of the basis, it is clear that the $\Delta_{\bm{q}}\sigma_{y}$ term pair electrons with net momentum $\bm{q}$ and opposite spin. Therefore, the Hamiltonian describes a superconductor which sustains a supercurrent when $\bm{q}$ is finite. With the Hamiltonian, we can calculate the spin magnetization $\bm{M}$ as ~\cite{Gorkov} 
\begin{align}\label{Magnetization_formula}
M_k=-\frac{1}{2\beta V}\sum_{\bm{p}, \bm{q}, n}\frac{\partial}{\partial h_k}\textrm{tr}\log G^{-1}\left(\bm{p}, \bm{q}, i\omega_n\right)|_{\bm{h}=0}.
\end{align}
Here $\beta$ reads $\beta^{-1}=k_bT$ with $k_b$ the Boltzmann constant, $k=x, y, z$ denotes the component of $\bm{M}$, and the Matsubara Green's function is defined as
\begin{align}
G^{-1}\left(\bm{p}, \bm{q}, i\omega_n\right)=i\omega_n-H\left(\bm{p}, \bm{q}\right)-\frac{1}{2}g\mu_{\textrm{B}}\begin{pmatrix}
\bm{h}\cdot\bm{\sigma} & 0 \\ 
0 & -\bm{h}\cdot\bm{\sigma}^\ast
\end{pmatrix},
\end{align}
with $\bm{h}$ describing a Zeeman field, $\mu_{\textrm{B}}$ the Bohr magneton, and $g$ the Lande $g$ factor. Since the supercurrent associated finite momentum $\bm{q}$ couples with the velocity operator as $\nabla\left(\xi_{\bm{p}}+\bm{g}_{\bm{p}}\cdot\bm{\sigma}\right)\cdot\frac{1}{2}\bm{q}$ at the linear order, the spin magnetic moment is induced through the rank-two pseudotensor $J_{ij}\left(\bm{p}\right)=\partial_{p_j}g_i\left(\bm{p}\right)$ with $i, j = x, y, z$. As a result, in the presence of supercurrent, the coupling term $\sigma_iJ_{ij}\frac{1}{2}q_j$ acts as an effective Zeeman field that generates  the spin magnetization.

In order to obtain an analytical form for the supercurrent induced magnetization $\bm{M}$, we expand $\textrm{tr}\log G^{-1}\left(\bm{p}, \bm{q}, i\omega_n\right)$ in terms of the Zeeman field $\bm{h}$ and the pairing order parameters $\Delta_{\bm{q}}$ and $\Delta^\ast_{\bm{q}}$. Then by taking the derivative on the Zeeman field $\bm{h}$ at $\bm{h}=0$ and summing over $\bm{p}$, $\bm{q}$, the Matsubara frequencies $\omega_n$, we obtain the analytical form for the magnetization $M$, which can be written as $M_k=T_{kj}\tilde{J}^{\textrm{S}}_{j}$, where  $\bm{J}^{\textrm{S}}$ is the supercurrent density. We denote $T_{kj}$ as a component of the magnetoelectric pseudotensor $T$~\cite{SM} with the form:
\begin{widetext}
\begin{align}\label{Key}
T_{kj}&=\left\{\frac{\Delta_0^2}{\bm{g}^2_{\bm{p}_{\textrm{F}}}}f\left(\frac{|\bm{g}_{\bm{p}_{\textrm{F}}}|}{\pi k_bT_c}\right)\left\langle J_{kj}\left(\bm{p}_{\textrm{F}}\right)\right\rangle+\left[\frac{7\Delta_0^2\zeta\left(3\right)}{8\pi^2k_b^2T_c^2}-\frac{\Delta_0^2}{\bm{g}^2_{\bm{p}_{\textrm{F}}}}f\left(\frac{|\bm{g}_{\bm{p}_{\textrm{F}}}|}{\pi k_bT_c}\right)\right]\left\langle\hat{\bm{g}}_{\bm{p}_{\textrm{F}}k}\hat{\bm{g}}_{\bm{p}_{\textrm{F}}i}J_{ij}\left(\bm{p}_{\textrm{F}}\right)\right\rangle\right\}\frac{2\mu_{\textrm{B}}g}{3\sqrt{3}\xi}N\left(E_{\textrm{F}}\right).
\end{align}
\end{widetext}
Here,  $\bm{p}_{\textrm{F}}$ represents the Fermi momentum, $N\left(E_{\textrm{F}}\right)$ is the density of states at the Fermi level, $\Delta_0$ is the homogeneous pairing order parameter magnitude, $\xi$ is the superconducting coherence length, and $\tilde{\bm{J}}^{\textrm{S}}=\bm{J}^{\textrm{S}}/J^{\textrm{S}}_{\textrm{max}}$ is the supercurrent density normalized by the maximum supercurrent density $J^{\textrm{S}}_{\textrm{max}}=2e\Delta_0^2\frac{2\hbar}{3\sqrt{3}m^\ast\xi}$ with $m^\ast$ being the effective pairing mass~\cite{Tinkham}. We have $\left\langle\right\rangle=\int\frac{d\Omega_{\bm{p}_{\textrm{F}}}}{4\pi}$ as the angle average at the Fermi surface and define the function $f\left(\rho\right)$ as
\begin{align}
f\left(\rho\right)=\textrm{Re}\sum_{n=0}^\infty\left(\frac{1}{2n+1}-\frac{1}{2n+1+i\rho}\right).
\end{align}
The magnetoelectric susceptibility obtained in Eq. \ref{Key} applies for superconductors with a pair of spin splitted bands. For noncentrosymmetric superconductor with multiple bands, the total magnetoelectric susceptibility is the sum of contribution from all the paired Fermi pockets as is discussed in the Supplementary Materials~\cite{SM}. As the magnetoelectric pseudotensor $T$ is constructed from the SOC pseudovector $\bm{g}_{\bm{p}_{\textrm{F}}}$ at the Fermi surface, its general form is determined by the crystal point group symmetry as shown in the next section.

\subsection{Symmetry Analysis for Magnetoelectric Pseudotensor in Three Dimensions}
The SOC pseudovector $\bm{g}_{\bm{p}}$ under the point group operation $\hat{R}$ respects $\bm{g}\left(\bm{p}\right)=\det\left(\hat{R}\right)\hat{R}\bm{g}\left(\hat{R}^{-1}\bm{p}\right)$~\cite{Samokhin2, Smidman}. Hence inheriting from the pseudovector $\bm{g}_{\bm{p}}$, we show that the magnetoelectric pseudotensor $T_{kj}$ under the crystal symmetry is subject to the constraints~\cite{SM}
\begin{align}\label{Transformation}
T=\det\left(\hat{R}\right)\hat{R}T\hat{R}^{\textrm{T}},
\end{align}
where $\hat{R}$ is the orthogonal matrix of the point group transformation. The general derivation of Eq. \ref{Transformation} for all the noncentrosymmetric point groups is given in the Materials and Methods section. All the nonzero components of the magnetoelectric pseudotensor $T$ of the 18 gyrotropic point groups are listed in Table \ref{magnetoelectric_pseudotensor}.

\newcommand{\tabincell}[2]{\begin{tabular}{@{}#1@{}}#2\end{tabular}}
\begin{table}[ht]
\caption{List of Magnetoelectric pseudotensor $T$ for the noncentrosymmetric superconductors in the 18 gyrotropic point groups. $T_{ij}$ with $i, j=x, y, z$ are in general the elements in $T$. $T_{\parallel}$ denotes the elements $T_{xx}$, $T_{yy}$ as $T_{\parallel}=T_{xx}=\pm T_{yy}$  and $T_0$ denotes the diagonal elements in T and O point group when  $T_{xx}=T_{yy}=T_{zz}$. $T_{\textrm{d}}$ and $T^-$ denote the symmetric and anti-symmetric off diagonal elements in $T$ respectively.The principal axis is along the $z$-direction.} 
\centering 
\begin{tabular}{c c c c c c} 
\hline\hline 
Point group & $T_{ij}$ & Point group & $T_{ij}$\\ [0.5ex] 
\hline 
C$_1$ & $\begin{pmatrix}
T_{xx} & T_{xy} & T_{xz} \\ 
T_{yx} & T_{yy} & T_{yz} \\ 
T_{zx} & T_{zy} & T_{zz}
\end{pmatrix}$ & C$_2$ & $\begin{pmatrix}
T_{xx} & T_{xy} & 0 \\ 
T_{yx} & T_{yy} & 0 \\ 
0 & 0 & T_{zz}
\end{pmatrix}$  \\
C$_3$ & $\begin{pmatrix}
T_{\parallel} & -T^- & 0 \\ 
T^- & T_{\parallel} & 0 \\ 
0 & 0 & T_{zz}
\end{pmatrix}$ & C$_4$ & $\begin{pmatrix}
T_{\parallel} & -T^- & 0 \\ 
T^- & T_{\parallel} & 0 \\ 
0 & 0 & T_{zz}
\end{pmatrix}$ \\ 
C$_6$ & $\begin{pmatrix}
T_{\parallel} & -T^- & 0 \\ 
T^- & T_{\parallel} & 0 \\ 
0 & 0 & T_{zz}
\end{pmatrix}$ & C$_{1v}$ & $\begin{pmatrix}
0 & T_{xy} & 0 \\ 
T_{yx} & 0 & T_{yz} \\ 
0 & T_{zy} & 0
\end{pmatrix}$ \\
C$_{2v}$ & $\begin{pmatrix}
0 & T_{xy} & 0 \\ 
T_{yx} & 0 & 0 \\ 
0 & 0 & 0
\end{pmatrix}$ & C$_{3v}$ & $\begin{pmatrix}
0 & -T^- & 0 \\ 
T^- & 0 & 0 \\ 
0 & 0 & 0
\end{pmatrix}$ \\
C$_{4v}$ & $\begin{pmatrix}
0 & -T^- & 0 \\ 
T^- & 0 & 0 \\ 
0 & 0 & 0
\end{pmatrix}$ &  C$_{6v}$ & $\begin{pmatrix}
0 & -T^- & 0 \\ 
T^- & 0 & 0 \\ 
0 & 0 & 0
\end{pmatrix}$ \\
D$_{2d}$ & $\begin{pmatrix}
T_{\parallel} & 0 & 0 \\ 
0 & -T_{\parallel} & 0 \\ 
0 & 0 & 0
\end{pmatrix}$ & S$_4$ & $\begin{pmatrix}
T_{\parallel} & T_{\textrm{d}} & 0 \\ 
T_{\textrm{d}} & -T_{\parallel} & 0 \\ 
0 & 0 & 0
\end{pmatrix}$ \\
D$_{2}$ & $\begin{pmatrix}
T_{xx} & 0 & 0 \\ 
0 & T_{yy} & 0 \\ 
0 & 0 & T_{zz}
\end{pmatrix}$ & D$_3$ & $\begin{pmatrix}
T_{\parallel} & 0 & 0 \\ 
0 & T_{\parallel} & 0 \\ 
0 & 0 & T_{zz}
\end{pmatrix}$ \\
D$_4$ & $\begin{pmatrix}
T_{\parallel} & 0 & 0 \\ 
0 & T_{\parallel} & 0 \\ 
0 & 0 & T_{zz}
\end{pmatrix}$ & D$_6$ & $\begin{pmatrix}
T_{\parallel} & 0 & 0 \\ 
0 & T_{\parallel} & 0 \\ 
0 & 0 & T_{zz}
\end{pmatrix}$ \\
T & $\begin{pmatrix}
T_0 & 0 & 0 \\ 
0 & T_0 & 0 \\ 
0 & 0 & T_0
\end{pmatrix}$ & O & $\begin{pmatrix}
T_0 & 0 & 0 \\ 
0 & T_0 & 0 \\ 
0 & 0 & T_0
\end{pmatrix}$ \\[1ex] 
\hline 
\end{tabular}
\label{magnetoelectric_pseudotensor} 
\end{table}

\subsection{Unconventional Magnetoelectric Effects in Gyrotropic Superconductors}
With the general form of $T_{ij}$ in Table \ref{magnetoelectric_pseudotensor}, some unconventional and novel magnetoelectric responses of gyrotropic superconductors can be identified immediately. One particular interesting case is for materials with point group symmetry T and O. In this case, the tensor $T_{ij}$ is proportional to the identity matrix and implies that the induced magnetization is parallel to the supercurrent direction shown in Fig. \ref{figure1} (b), namely, $\bm{M} \propto\tilde{\bm{J}}^{\textrm{S}}$. This is a quantum analogue of classical solenoids but without the need to fabricate any helical structures and the longitudinal response is induced by SOC. It is interesting to note that a few superconductors with relatively high $T_c$ have point group O such Li$_2$Pt$_3$B, Li$_2$Pd$_3$B and Mo$_3$Al$_2$C~\cite{Salamon, Hirata, Prozorov}. Another interesting examples are superconducting TaRh$_2$B$_2$ and NbRh$_2$B$_2$ which had been newly discovered by Carnicom et al.~\cite{Carnicom}. These superconductors have point group C$_3$. From Table \ref{magnetoelectric_pseudotensor}, we immediately realize that a supercurrent along the $z$-direction (the principal symmetry axis direction) will generate a pure longitudinal response such that the supercurrent-induced magnetization is parallel to the direction of the supercurrent. On the other hand, a supercurrent perpendicular to the $z$-direction generates a magnetization in the $x$-$y$ plane.

Importantly, in the normal state, these materials with chiral lattice symmetries are Kramers Weyl Semimetals which has Weyl points pinned at time-reversal invariant momenta~\cite{Guoqing}. Therefore, using Table \ref{magnetoelectric_pseudotensor}, one can immediately identify the novel magnetoelectric properties of a superconducting Kramers Weyl semimetal. More noncentrosymmetric superconductors with gyrotropic point groups are listed in the Supplementary Materials~\cite{SM}. 

Another interesting result from Table \ref{magnetoelectric_pseudotensor} is that, in quasi-two-dimensions, only C$_1$, C$_{1v}$, C$_{2v}$ and C$_2$ symmetries with polar axis lying inside the atomic plane allow an out-of-plane magnetization induced by an in-plane supercurrent, as is present in Fig. \ref{figure1} (c). Interestingly, recently discovered few layer 1Td-WTe$_2$~\cite{Sajadi, Sanfeng} and 1Td-MoTe$_2$~\cite{Guangtong} have such a low symmetry: C$_{1v}$. Moreover, due to the large SOC in these materials, the magnetoelectric effect is expected to be strong. With C$_{1v}$ symmetry, we show below how the magnetization direction can also be controlled by the direction of the supercurrent. The magnetoelectric effect of 2H-structure TMDs are also discussed in the next section.

\begin{figure}
\centering
\includegraphics[width=3.6in]{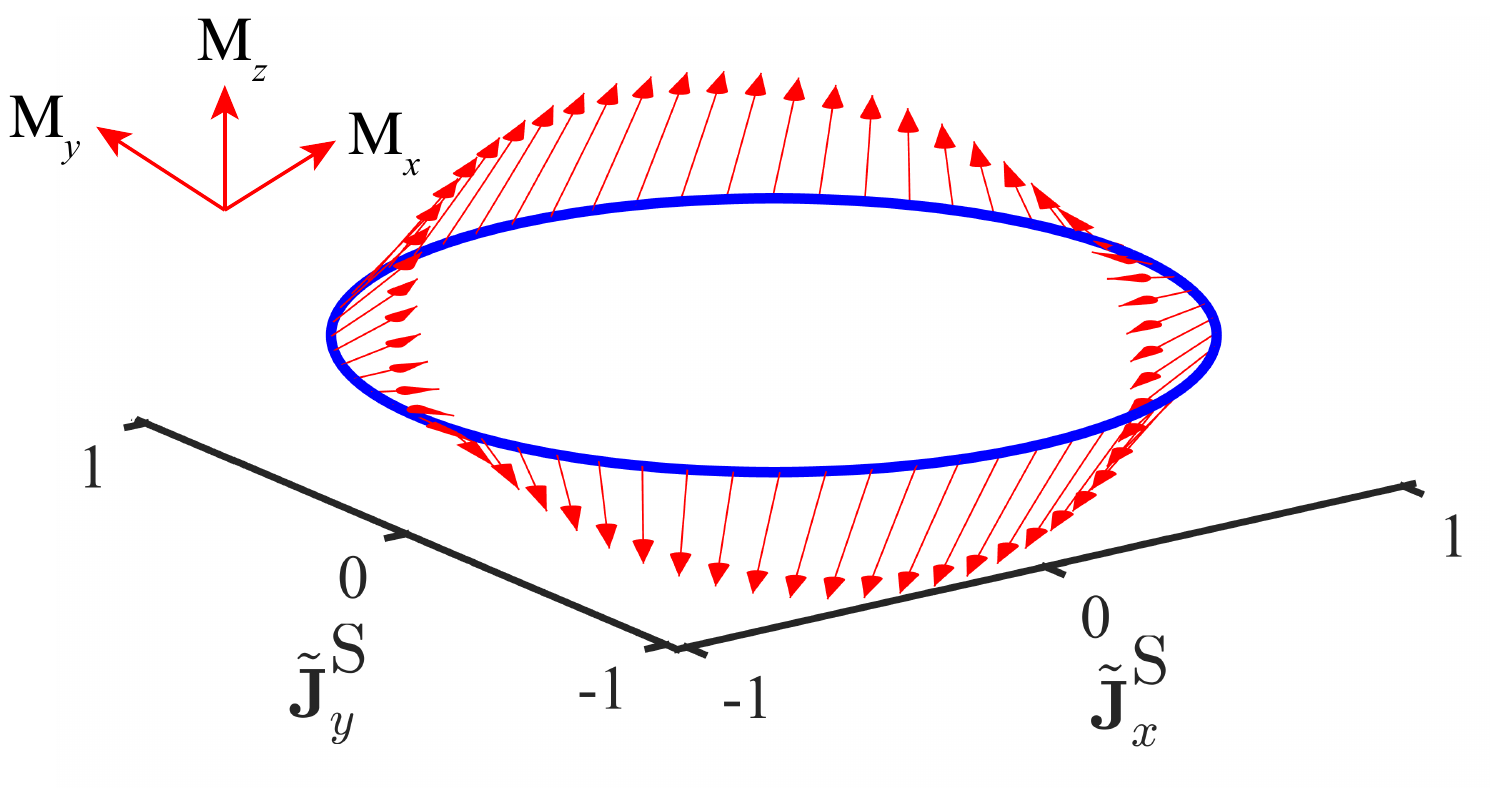}
\caption{The magnetization direction (red arrows) can be controlled by the in-plane supercurrent directions $\tilde{\bm{J}}^{\textrm{S}}$ as calculated using realistic parameters for bilayer 1Td-MoTe$_2$. The in-plane mirror invariant line is defined along the $\tilde{\bm{J}}_x^{\textrm{S}}$ direction.}\label{figure2}
\end{figure}

\subsection{Unconventional magnetoelectric effects of superconducting TMDs}
In this section, we apply the general theory obtained above to two interesting TMDs, namely, bilayer or multilayer 1Td-MoTe$_2$ and strained monolayer 2H-NbSe$_2$. For the case of bilayer 1Td-MoTe$_2$, the crystal structure belongs to the C$_{1v}$ point group and only has one in-plane mirror symmetry with $y\rightarrow-y$. The superconductivity occurs at the critical temperature $T_c=4$K and it has the in-plane anisotropic upper critical field $H_{c2}$ which exceeds the Pauli limit $H_p$ due to its anisotropic SOC $\bm{g}\left(p_x, p_y\right)=\left(v_xp_y, v_yp_x, v_zp_y\right)$. From the angle dependent in-plane $H_{c2}\left(\phi\right)$, the anisotropic SOC strength at the Fermi level is estimated to be $\bm{g}\left(\phi\right)=\left(16.9\sin\phi, 23.5\cos\phi, 23.1\sin\phi\right)$meV~\cite{Guangtong}. With the Fermi wave vector $\left(p_{\textrm{F}x}, p_{\textrm{F}y}\right)=\left(1.009, 1.050\right)\textrm{nm}^{-1}$ and the Fermi energy density of states $N\left(E_{\textrm{F}}\right)=4.3776\textrm{eV}^{-1}\textrm{nm}^{-2}$ from the first principle calculation~\cite{Guangtong}, we can further use the pairing gap $\Delta_0=1.76k_bT_c$ at zero temperature and the coherence length $\xi=20.79$nm ~\cite{Guangtong} to estimate the magnetoelectric pseudotensor $T$ as $T_{xy}=2.52\times10^{-4}\mu_{\textrm{B}}/\textrm{nm}^2$, $T_{yx}=2.64\times10^{-4}\mu_{\textrm{B}}/\textrm{nm}^2$ and $T_{zy}=3.43\times10^{-3}\mu_{\textrm{B}}/\textrm{nm}^2$ while all other elements are zero. 

From the form of $T$, we note that the supercurrent induced magnetization $\left(M_x, M_y, M_z\right)$ is a function of the in-plane supercurrent direction shown in Fig.\ref{figure2}. When the supercurrent flows along $y$ direction, the $z$ direction magnetization $M_z$ gets the optimised value. For a supercurrent close to the critical value, the maximum $M_z$ can reach $M_z=7.6\times10^{-4}\mu_{\textrm{B}}$ per unit cell. The strength of supercurrent induced magnetization is in the same order as Bi/Ag bilayers with strong Rashba SOC~\cite{Johansson1, Johansson2} which has been observed \cite{Fert}. Therefore, we expect that 1Td-MoTe$_2$ is an excellent candidate to observe the superconducting magnetoelectric effect which has yet to be observed experimentally. Moreover, our prediction of the supercurrent induced magnetization also applies to the recently realized superconducting 1Td-WTe$_2$~\cite{Sajadi, Sanfeng}.

For the case of 2H-NbSe$_2$ which is also known as an Ising superconductor~\cite{Jianting, Saito, Mak, Benjamin, Wenyu}, the materials have point group symmetry D$_{3h}$~\cite{Noah}.  Due to the in-plane mirror invariant line at $x=0$ and the three-fold rotation symmetry, the SOC takes the form $g_z\left(\bm{p}\right)\propto p_x\left(p_x^2-3p_y^2\right)$ around the $\Gamma$ pocket and $g_z\left(\bm{p}\mp K\right)=\pm\beta_{\textrm{so}}$ around the $K\left(K'\right)$ pockets\cite{Wenyu}. According to the general analysis above, the three-fold rotation symmetry forces the magnetoelectric pseudotensor $T$ to be zero and there is no magnetoelectric effect. This is an interesting example of a noncentrosymmetric superconductors with zero magnetoelectric response. Interestingly, a uniaxial strain that breaks the three-fold symmetry and reduce the point group symmetry to C$_{2v}$ with the polar axis lying inside the atomic plane. As a result, SOC which is linearly proportional to $p_x$, namely, $g_z\left(\bm{p}\right)\propto p_x$, can be induced. Therefore, under uniaxial strain, the magnetoelectric pseudotensor $T$ will have nonzero element $T_{zx}$ and the $x$ component of the supercurrent will generate the out-of-plane magnetization.  This provides a novel way to use strain to generate and manipulate spin polarizations in non-gyrotropic superconductors. 

\section{Discussion}
In this work, we presented the general form of the magnetoelectric pseudotensor for gyrotropic superconductors as summerized in Table \ref{magnetoelectric_pseudotensor}. Our theory provides a powerful tool for the search of unconventional magnetoelectric effects (or the Edelstein effect) in materials. Guided by the general theory, we demonstrated novel ways of generating and manipulating spin polarizations in noncentrosymmetric superconductors such as 1Td-MoTe$_2$ and 2H-NbSe$_2$. In particularly, we predict that 1Td-MoTe$_2$ is an excellent candidate for the first realization of the Edelstein effect in superconductors.

Importantly, Table \ref{magnetoelectric_pseudotensor} can be used to identify the magnetoelectric effects of a large number of superconductors with chiral lattice symmetry such as Li$_2$Pt$_3$B, Li$_2$Pd$_3$B and Mo$_3$Al$_2$C with O point group \cite{Salamon, Hirata, Prozorov}, and newly discovered superconducting chiral crystals such as TaRh$_2$B$_2$ and NbRh$_2$B$_2$ with C$_{3}$ point group~\cite{Carnicom}. The supercurrent driven longitudinal magnetization in those chiral crystals resembles the current-induced magnetization of solenoids at microscopic scale. These superconducting chiral crystals can have potential applications for new designs superconducting spintronic devices~\cite{Linder}. 

It is also important to note that the Edelstein effect (current-induced spin magnetization effect) had been observed in several materials with Rashba SOC in the normal state~\cite{Gambardella}. However, Edelstein effect has not been observed in superconducting materials. In this work, we identified several superconducting materials, such as 1Td-MoTe$_2$ and strained NbSe$_2$, which possess strong Edelstein effect and the proposed Edelstein effect can be observed through magneto-optical Kerr effect measurements~\cite{Fai}. 

\section*{Acknowledgement}
W.-Y. He and K. T. Law are thankful for the support of HKRGC through C6026-16W, 16324216, 16307117 and 16309718. K. T. Law is further supported by the Croucher Foundation and the Dr. Tai-chin Lo Foundation.


\onecolumngrid
\clearpage
\begin{center}
{\bf Supplementary Material for ``Magnetoelectric Effects in Gyrotropic Superconductors''}
\end{center}

\author{Wen-Yu He and K. T. Law}\thanks{Correspondence address: phlaw@ust.hk}
\affiliation{Department of Physics, Hong Kong University of Science and Technology, Clear Water Bay, Hong Kong, China}

\maketitle
\setcounter{equation}{0}
\setcounter{figure}{0}
\setcounter{table}{0}
\setcounter{page}{1}
\makeatletter
\renewcommand{\theequation}{S\arabic{equation}}
\renewcommand{\thefigure}{S\arabic{figure}}
\renewcommand{\thetable}{S\arabic{table}}
\renewcommand{\bibnumfmt}[1]{[S#1]}
\renewcommand{\citenumfont}[1]{S#1}

\section{General theory for the magnetoelectric Effect}
In the singlet pairing channel, we consider the on-site pairing interaction and write the saddle point approximated partition function for the superconductor with spin-orbit coupling and Zeeman field as
\begin{align}
Z\left(\bm{h}\right)\approx e^{-\beta V\left[\sum_{\bm{q}}\frac{\Delta_{\bm{q}}^\ast\Delta_{\bm{q}}}{U_0}-\frac{1}{2\beta V}\sum_{\bm{p}, \bm{q}, n}\textrm{tr}\log G^{-1}\left(\bm{p}, \bm{q}, i\omega_n\right)\right]},
\end{align}
where $U_0$ is amplitude for the singlet pairing interaction and the Green's function reads
\begin{align}
G^{-1}\left(\bm{p}, \bm{q}, i\omega_n\right)=\begin{pmatrix}
G^{-1}_e\left(\bm{p}, \bm{q}, i\omega_n\right)-\frac{1}{2}\mu_{\textrm{B}}g\bm{h}\cdot\bm{\sigma} & -i\Delta_{\bm{q}}\sigma_y\\ 
i\Delta^\ast_{\bm{q}}\sigma_y & G^{-1}_h\left(\bm{p}, \bm{q}, i\omega_n\right)+\frac{1}{2}\mu_{\textrm{B}}g\bm{h}\cdot\bm{\sigma}^\ast
\end{pmatrix},
\end{align}
with $\mu_{\textrm{B}}$ being the Bohr magneton and $g$ being the Lande $g$ factor. The Green's functions for the electron and hole are defined as
\begin{align}
G^{-1}_{\textrm{e}}\left(\bm{p}, \bm{q}, i\omega_n\right)&=i\omega_n-\xi_{\bm{p}+\frac{1}{2}\bm{q}}-\bm{g}_{\bm{p}+\frac{1}{2}\bm{q}}\cdot\bm{\sigma}\approx i\omega_n-\xi_{\bm{p}}-\bm{g}_{\bm{p}}\cdot\bm{\sigma}-\nabla\left(\xi_{\bm{p}}+\bm{g}_{\bm{p}}\cdot\bm{\sigma}\right)\cdot\frac{1}{2}\bm{q},\\
G^{-1}_{\textrm{h}}\left(\bm{p}, \bm{q}, i\omega_n\right)&=i\omega_n+\xi_{-\bm{p}+\frac{1}{2}\bm{q}}+\bm{g}_{-\bm{p}+\frac{1}{2}\bm{q}}\cdot\bm{\sigma}^\ast\approx i\omega_n+\xi_{\bm{p}}-\bm{g}_{\bm{p}}\cdot\bm{\sigma}^\ast-\nabla\left(\xi_{\bm{p}}-\bm{g}_{\bm{p}}\cdot\bm{\sigma}^\ast\right)\cdot\frac{1}{2}\bm{q},
\end{align}
where $\bm{g}_{\bm{p}}$ is the spin-orbit coupling field satisfying $\bm{g}_{\bm{p}}=-\bm{g}_{-\bm{p}}$. We then use the Dyson's equation to simplify the Green's functions to the linear order in $\bm{q}$ as
\begin{align}\label{Ge1}
G_{\textrm{e}}\left(\bm{p}, \bm{q}, i\omega_n\right)&\approx G_{\textrm{e}}\left(\bm{p}, i\omega_n\right)+G_{\textrm{e}}\left(\bm{p}, i\omega_n\right)\bm{v}_{\textrm{e}}\cdot\frac{1}{2}\bm{q}G_{\textrm{e}}\left(\bm{p}, i\omega_n\right)+\mathcal{O}\left(\bm{q}^2\right),\\\label{Gh1}
G_{\textrm{h}}\left(\bm{p}, \bm{q}, i\omega_n\right)&\approx G_h\left(\bm{p}, i\omega_n\right)+G_h\left(\bm{p}, i\omega_n\right)\bm{v}_{\textrm{h}}\cdot\frac{1}{2}\bm{q}G_{\textrm{h}}\left(\bm{p}, i\omega_n\right)+\mathcal{O}\left(\bm{q}^2\right),
\end{align}
where the velocity operator takes the form
\begin{align}
\bm{v}_{\textrm{e}}=\nabla\left(\xi_{\bm{p}}+\bm{g}_{\bm{p}}\cdot\bm{\sigma}\right),\quad\quad\quad\bm{v}_{\textrm{h}}=\nabla\left(\xi_{\bm{p}}-\bm{g}_{\bm{p}}\cdot\bm{\sigma}^\ast\right).
\end{align}
The Green's function $G_{\textrm{e}}\left(\bm{p}, i\omega_n\right)$ and $G_{\textrm{h}}\left(\bm{p}, i\omega_n\right)$ are further expressed as
\begin{align}\label{Geh1}
G_{\textrm{e}}\left(\bm{p}, i\omega_n\right)=G^+_{\textrm{e}}\left(\bm{p}, i\omega_n\right)+G^-_{\textrm{e}}\left(\bm{p}, i\omega_n\right)\hat{\bm{g}}_{\bm{p}}\cdot\bm{\sigma},\quad G_{\textrm{h}}\left(\bm{p}, i\omega_n\right)=G^+_{\textrm{h}}\left(\bm{p}, i\omega_n\right)+G^-_{\textrm{h}}\left(\bm{p}, i\omega_n\right)\hat{\bm{g}}_{\bm{p}}\cdot\bm{\sigma}^\ast,
\end{align}
with $\hat{\bm{g}}_{\bm{p}}=\frac{\bm{g}_{\bm{p}}}{|\bm{g}_{\bm{p}}|}$ and
\begin{align}\label{Geh2}
G^{\pm}_{\textrm{e}}\left(\bm{p}, i\omega_n\right)=\frac{1}{2}\left[\frac{1}{i\omega_n-\xi_{\bm{p}}-|\bm{g}_{\bm{p}}|}\pm\frac{1}{i\omega_n-\xi_{\bm{p}}+|\bm{g}_{\bm{p}}|}\right],\quad G^{\pm}_{\textrm{h}}=\frac{1}{2}\left[\frac{1}{i\omega_n+\xi_{\bm{p}}-|\bm{g}_{\bm{p}}|}\pm\frac{1}{i\omega_n+\xi_{\bm{p}}+|\bm{g}_{\bm{p}}|}\right].
\end{align}

The magnetization $\bm{M}$ can be obtained from the partition function as
\begin{align}\nonumber
\bm{M}&=-\frac{1}{\beta V}\frac{\partial}{\partial \bm{h}}\log Z\left(\bm{h}\right)\\
&=-\frac{1}{2\beta V}\sum_{\bm{p}, \bm{q}, n}\frac{\partial}{\partial\bm{h}}\textrm{tr}\log G^{-1}\left(\bm{p}, \bm{q}, i\omega_n\right).
\end{align}
We decompose the Green's function $G^{-1}\left(\bm{p}, \bm{q}, i\omega_n\right)=G^{-1}_0\left(\bm{p}, \bm{q}, i\omega_n\right)-\Sigma$ with
\begin{align}
G^{-1}_0\left(\bm{p}, \bm{q}, i\omega_n\right)=\begin{pmatrix}
G^{-1}_{\textrm{e}}\left(\bm{p}, \bm{q}, i\omega_n\right) & -i\Delta_{\bm{q}}\sigma_y\\
i\Delta^\ast_{\bm{q}}\sigma_y  & G^{-1}_h\left(\bm{p}, \bm{q}, i\omega_n\right)
\end{pmatrix},\quad\Sigma=\frac{1}{2}\mu_{\textrm{B}}g\begin{pmatrix}
\bm{h}\cdot\bm{\sigma} & 0\\
0 & -\bm{h}\cdot\bm{\sigma}^\ast
\end{pmatrix}.
\end{align}
We can then expand the logarithm in powers of the self-energy $\Sigma$
\begin{align}
\log G^{-1}\left(\bm{p}, \bm{q}, i\omega_n\right)=\log G^{-1}_0\left(\bm{p}, \bm{q}, i\omega_n\right)-\sum_{l=1}^{\infty}\frac{1}{l}\left[G_0\left(\bm{p}, \bm{q}, i\omega_n\right)\Sigma\right]^l.
\end{align}
We keep the first order correction and then the magnetization $\bm{M}$ becomes
\begin{align}
\bm{M}\approx\frac{1}{2\beta V}\sum_{\bm{p}, \bm{q}, n}\frac{\partial}{\partial\bm{h}}\textrm{tr}\left[G_0\left(\bm{p}, \bm{q}, i\omega_n\right)\Sigma\right].
\end{align}
We then further decompose the Green's function $G^{-1}_0\left(\bm{p}, \bm{q}, i\omega_n\right)=G^{-1}_{\textrm{N}}-\Sigma_{\Delta}$ with
\begin{align}
G^{-1}_{\textrm{N}}\left(\bm{p}, \bm{q}, i\omega_n\right)=\begin{pmatrix}
G^{-1}_{\textrm{e}}\left(\bm{p}, \bm{q}, i\omega_n\right) & 0\\ 
0 & G^{-1}_{\textrm{h}}\left(\bm{p}, \bm{q}, i\omega_n\right)
\end{pmatrix},\quad\Sigma_{\Delta}=\begin{pmatrix}
0 & i\Delta_{\bm{q}}\sigma_y\\ 
-i\Delta^\ast_{\bm{q}}\sigma_y & 0
\end{pmatrix},
\end{align}
and expand the Green's function $G_0\left(\bm{p}, \bm{q}, i\omega_n\right)$ in power series as
\begin{align}\nonumber
G_0\left(\bm{p}, \bm{q}, i\omega_n\right)&\approx G_{\textrm{N}}\left(\bm{p}, \bm{q}, i\omega_n\right)+G_{\textrm{N}}\left(\bm{p}, \bm{q}, i\omega_n\right)\Sigma_{\Delta}G_{\textrm{N}}\left(\bm{p}, \bm{q}, i\omega_n\right)\\
&+G_{\textrm{N}}\left(\bm{p}, \bm{q}, i\omega_n\right)\Sigma_{\Delta}G_{\textrm{N}}\left(\bm{p}, \bm{q}, i\omega_n\right)\Sigma_{\Delta}G_{\textrm{N}}\left(\bm{p}, \bm{q}, i\omega_n\right)+...
\end{align}
As a result, the magnetization $\bm{M}$ is approximated as
\begin{align}\nonumber
\bm{M}&\approx\frac{1}{2\beta V}\sum_{\bm{p}, \bm{q}, n}\frac{\partial}{\partial\bm{h}}\textrm{tr}\left[G_{\textrm{N}}\left(\bm{p}, \bm{q}, i\omega_n\right)\Sigma+G_{\textrm{N}}\left(\bm{p}, \bm{q}, i\omega_n\right)\Sigma_{\Delta}G_{\textrm{N}}\left(\bm{p}, \bm{q}, i\omega_n\right)\Sigma\right.\\
&\left.+G_{\textrm{N}}\left(\bm{p}, \bm{q}, i\omega_n\right)\Sigma_{\Delta}G_{\textrm{N}}\left(\bm{p}, \bm{q}, i\omega_n\right)\Sigma_{\Delta}G_{\textrm{N}}\left(\bm{p}, \bm{q}, i\omega_n\right)\Sigma\right].
\end{align}
We find that the first term is responsible for the magnetoelectric effect in the normal metal phase and the second term will be zero after the trace and the summation, so the magnetization $\bm{M}$ is simplified as
\begin{align}\nonumber\label{M1}
\bm{M}&=\frac{\mu_{\textrm{B}}g}{4\beta V}\sum_{\bm{p}, \bm{q}, n}\frac{\partial}{\partial\bm{h}}\Delta^\ast_{\bm{q}}\Delta_{\bm{q}}\textrm{tr}\left[G_{\textrm{e}}\left(\bm{p}, \bm{q}, i\omega_n\right)\sigma_yG_{\textrm{h}}\left(\bm{p}, \bm{q}, i\omega_n\right)\sigma_yG_{\textrm{e}}\left(\bm{p}, \bm{q}, i\omega_n\right)\bm{h}\cdot\bm{\sigma}\right.\\
&\left.-G_{\textrm{h}}\left(\bm{p}, \bm{q}, i\omega_n\right)\sigma_yG_{\textrm{e}}\left(\bm{p}, \bm{q}, i\omega_n\right)\sigma_yG_{\textrm{h}}\left(\bm{p}, \bm{q}, i\omega_n\right)\bm{h}\cdot\bm{\sigma}^\ast\right].
\end{align}
We then substitute Eq. \ref{Ge1} and Eq. \ref{Gh1} into Eq. \ref{M1}. With the velocity operator $\bm{v}_{\textrm{e}/\textrm{h}}$ and Eq. \ref{Geh1}, Eq. \ref{Geh2}, the component of the magnetization $\bm{M}$ becomes
\begin{align}\nonumber
M_k&=\frac{\mu_{\textrm{B}}g}{4\beta V}\sum_{\bm{p}, \bm{q}, n}\Delta^\ast_{\bm{q}}\Delta_{\bm{q}}\textrm{tr}\left\{\left[G_e^+\left(G_e^{+2}G_h^++G_e^{-2}G_h^+-2G_e^+G_e^-G_h^-\right)+\left(G_e^+G_h^+-G_e^-G_h^-\right)^2\right.\right.\\\nonumber
&\left.\left.+G_h^+\left(G_h^{+2}G_e^++G_h^{-2}G_e^+-2G_h^+G_h^-G_e^-\right)\right]\nabla\left(\bm{g}_{\bm{p}}\cdot\bm{\sigma}\right)\cdot\bm{q}\sigma_k+\left[-G_e^-\left(G_e^{-2}G_h^-+G_e^{+2}G_h^--2G_e^+G_e^-G_h^+\right)\right.\right.\\
&\left.\left.+\left(G_e^+G_h^--G_e^-G_h^+\right)^2-G_h^-\left(G_h^{+2}G_e^-+G_h^{-2}G_e^--2G_h^+G_h^-G_e^+\right)\right]\left(\hat{\bm{g}}_{\bm{p}}\cdot\bm{\sigma}\right)\nabla\left(\bm{g}_{\bm{\sigma}}\cdot\bm{\sigma}\right)\cdot\bm{q}\left(\hat{\bm{g}}_{\bm{p}}\cdot\bm{\sigma}\right)\sigma_k\right\}.
\end{align}
As the gradient of the spin-orbit coupling pseudovector function $\bm{g}\left(\bm{p}\right)$ is a rank-two pseudotensor $J_{ij}\left(\bm{p}\right)=\frac{\partial g_i\left(\bm{p}\right)}{\partial p_j}$ with the matrix form
\begin{align}
\bm{J}\left(\bm{p}\right)=\begin{pmatrix}
\partial_{p_x} g_x\left(\bm{p}\right) & \partial_{p_y} g_x\left(\bm{p}\right) & \partial_{p_z} g_x\left(\bm{p}\right)\\ 
\partial_{p_x} g_y\left(\bm{p}\right) & \partial_{p_y} g_y\left(\bm{p}\right) & \partial_{p_z} g_y\left(\bm{p}\right)\\ 
\partial_{p_x} g_z\left(\bm{p}\right) & \partial_{p_y} g_z\left(\bm{p}\right) & \partial_{p_z} g_z\left(\bm{p}\right)
\end{pmatrix},
\end{align}
then the term $\nabla\left(\bm{g}_{\bm{p}}\cdot\bm{\sigma}\right)\cdot\bm{q}$ is expressed as $\nabla\left(\bm{g}_{\bm{p}}\cdot\bm{\sigma}\right)\cdot\bm{q}=\sigma_iJ_{ij}\left(\bm{p}\right)q_j$. As a result, the magnetization $\bm{M}$ becomes
\begin{align}\nonumber
M_k&=\frac{\mu_{\textrm{B}}g}{4\beta V}\sum_{\bm{p}, q_j, n}\Delta^\ast_{q_j}\Delta_{q_j}\left\{\left[G_e^+\left(G_e^{+2}G_h^++G_e^{-2}G_h^+-2G_e^+G_e^-G_h^-\right)+\left(G_e^+G_h^+-G_e^-G_h^-\right)^2\right.\right.\\\nonumber
&\left.\left.+G_h^+\left(G_h^{+2}G_e^++G_h^{-2}G_e^+-2G_h^+G_h^-G_e^-\right)\right]\textrm{tr}\left[\sigma_iJ_{ij}\left(\bm{p}\right)q_j\sigma_k\right]+\left[-G_e^-\left(G_e^{-2}G_h^-+G_e^{+2}G_h^--2G_e^+G_e^-G_h^+\right)\right.\right.\\
&\left.\left.+\left(G_e^+G_h^--G_e^-G_h^+\right)^2-G_h^-\left(G_h^{+2}G_e^-+G_h^{-2}G_e^--2G_h^+G_h^-G_e^+\right)\right]\textrm{tr}\left[\hat{g}_{\bm{p}m}\sigma_m\sigma_iJ_{ij}\left(\bm{p}\right)q_j\hat{g}_{\bm{p}l}\sigma_l\sigma_k\right]\right\}.
\end{align}
We notice that the trace for the Pauli matrix has the relations
\begin{align}
\textrm{tr}\left(\sigma_a\sigma_b\right)=2\delta_{ab},\quad\quad\textrm{tr}\left(\sigma_a\sigma_b\sigma_c\right)=2i\epsilon_{abc},\quad\quad\textrm{tr}\left(\sigma_a\sigma_b\sigma_c\sigma_d\right)=2\left(\delta_{ab}\delta_{cd}-\delta_{ac}\delta_{bd}+\delta_{ad}\delta_{bc}\right).
\end{align}
Then the magnetization $\bm{M}$ is written as
\begin{align}\nonumber\label{Magnetization}
M_k&=\frac{\mu_{\textrm{B}}g}{2\beta V}\sum_{q_j}q_j\Delta^\ast_{q_j}\Delta_{q_j}\sum_{\bm{p}, n}\left\{\left[G_e^+\left(G_e^{+2}G_h^++G_e^{-2}G_h^+-2G_e^+G_e^-G_h^-\right)+\left(G_e^+G_h^+-G_e^-G_h^-\right)^2\right.\right.\\\nonumber
&\left.\left.+G_h^+\left(G_h^{+2}G_e^++G_h^{-2}G_e^+-2G_h^+G_h^-G_e^-\right)\right]J_{kj}\left(\bm{p}\right)+\left[-G_e^-\left(G_e^{-2}G_h^-+G_e^{+2}G_h^--2G_e^+G_e^-G_h^+\right)\right.\right.\\\nonumber
&\left.\left.+\left(G_e^+G_h^--G_e^-G_h^+\right)^2-G_h^-\left(G_h^{+2}G_e^-+G_e^{-2}G_e^--2G_h^+G_h^-G_e^+\right)\right]\left[\hat{g}_{\bm{p}i}J_{ij}\left(\bm{p}\right)\hat{g}_{\bm{p}k}-\hat{g}_{\bm{p}l}J_{kj}\left(\bm{p}\right)\hat{g}_{\bm{p}l}+\hat{g}_{\bm{p}k}J_{lj}\left(\bm{p}\right)\hat{g}_{\bm{p}l}\right]\right\}\\\nonumber
&=\frac{\mu_{\textrm{B}}g}{2\beta V}\sum_{q_j}q_j\Delta^\ast_{q_j}\Delta_{q_j}\sum_{\bm{p}, n}\left\{\left[G_e^+\left(G_e^{+2}G_h^++G_e^{-2}G_h^+-2G_e^+G_e^-G_h^-\right)+\left(G_e^+G_h^+-G_e^-G_h^-\right)^2\right.\right.\\\nonumber
&\left.\left.+G_h^+\left(G_h^{+2}G_e^++G_h^{-2}G_e^+-2G_h^+G_h^-G_e^-\right)+G_e^-\left(G_e^{-2}G_h^-+G_e^{+2}G_h^--2G_e^+G_e^-G_h^+\right)\right.\right.\\\nonumber
&\left.\left.-\left(G_e^+G_h^--G_e^-G_h^+\right)^2+G_h^-\left(G_h^{+2}G_e^-+G_h^{-2}G_e^--2G_h^+G_h^-G_e^+\right)\right]J_{kj}\left(\bm{p}\right)\right.\\\nonumber
&\left.-2\left[G_e^-\left(G_e^{-2}G_h^-+G_e^{+2}G_h^--2G_e^+G_e^-G_h^+\right)-\left(G_e^+G_h^--G_e^-G_h^+\right)^2\right.\right.\\
&\left.\left.+G_h^-\left(G_h^{+2}G_e^-+G_h^{-2}G_e^--2G_h^+G_h^-G_e^+\right)\right]\hat{g}_{\bm{p}i}J_{ij}\left(\bm{p}\right)\hat{g}_{\bm{p}k}\right\}.
\end{align}
For the integral of the Green's function series, we have
\begin{align}\nonumber\label{Green1}
&\frac{k_bT}{V}\sum_{\bm{p}, n}\left[G_e^+\left(G_e^{+2}G_h^++G_e^{-2}G_h^+-2G_e^+G_e^-G_h^-\right)+\left(G_e^+G_h^+-G_e^-G_h^-\right)^2+G_h^+\left(G_h^{+2}G_e^++G_h^{-2}G_e^+-2G_h^+G_h^-G_e^-\right)\right.\\\nonumber
+&\left.G_e^-\left(G_e^{-2}G_h^-+G_e^{+2}G_h^--2G_e^+G_e^-G_h^+\right)-\left(G_e^+G_h^--G_e^-G_h^+\right)^2+G_h^-\left(G_h^{+2}G_e^-+G_h^{-2}G_e^--2G_h^+G_h^-G_e^+\right)\right]J_{kj}\left(\bm{p}\right)\\\nonumber
=&N\left(E_{\textrm{F}}\right)k_bT\sum_n\int d\xi\left\{\frac{1}{2|\bm{g}_{\bm{p}_{\textrm{F}}}|}\frac{\xi+|\bm{g}_{\bm{p}_{\textrm{F}}}|}{\left[\omega_n^2+\left(\xi+|\bm{g}_{\bm{p}_{\textrm{F}}}|\right)^2\right]^2}-\frac{1}{2|\bm{g}_{\bm{p}_{\textrm{F}}}|}\frac{\xi-|\bm{g}_{\bm{p}_{\textrm{F}}}|}{\left[\omega_n^2+\left(\xi-|\bm{g}_{\bm{p}_{\textrm{F}}}|\right)^2\right]^2}\right.\\\nonumber
+&\left.\frac{2}{\left[\omega_n^2+\left(\xi+|\bm{g}_{\bm{p}_{\textrm{F}}}|\right)^2\right]\left[\omega_n^2+\left(\xi-|\bm{g}_{\bm{p}_{\textrm{F}}}|\right)^2\right]}\right\}\int\frac{d\Omega}{4\pi}J_{kj}\left(\bm{p}_{\textrm{F}}\right)\\
=&2N\left(E_{\textrm{F}}\right)k_bT\sum_{n=0}^\infty\frac{\pi}{\omega_n\left(\omega_n^2+|\bm{g}_{\bm{p}_{\textrm{F}}}|^2\right)}\left\langle J_{kj}\left(\bm{p}_{\textrm{F}}\right)\right\rangle,
\end{align}
and
\begin{align}\nonumber\label{Green2}
&-\frac{k_bT}{V}\sum_{\bm{p}, n}2\left[G_e^-\left(G_e^{-2}G_h^-+G_e^{+2}G_h^--2G_e^+G_e^-G_h^+\right)-\left(G_e^+G_h^--G_e^-G_h^+\right)^2\right.\\\nonumber
&\left.+G_h^-\left(G_h^{+2}G_e^-+G_h^{-2}G_e^--2G_h^+G_h^-G_e^+\right)\right]\hat{g}_{\bm{p}i}J_{ij}\left(\bm{p}\right)\hat{g}_{\bm{p}k}\\\nonumber
=&-N\left(E_{\textrm{F}}\right)k_bT\sum_n\int d\xi\left\{\frac{2\left(\xi+|\bm{g}_{\bm{p}_{\textrm{F}}}|\right)^2}{\left[\omega_n^2+\left(\xi+|\bm{g}_{\bm{p}_{\textrm{F}}}|\right)^2\right]^3}+\frac{2\left(\xi-|\bm{g}_{\bm{p}_{\textrm{F}}}|\right)^2}{\left[\omega_n^2+\left(\xi-|\bm{g}_{\bm{p}_{\textrm{F}}}|\right)^2\right]^3}+\frac{\xi+|\bm{g}_{\bm{p}_{\textrm{F}}}|}{2|\bm{g}_{\bm{p}_{\textrm{F}}}|\left[\omega_n^2+\left(\xi+|\bm{g}_{\bm{p}_{\textrm{F}}}|\right)^2\right]^2}\right.\\\nonumber
&\left.-\frac{\xi-|\bm{g}_{\bm{p}_{\textrm{F}}}|}{2|\bm{g}_{\bm{p}_{\textrm{F}}}|\left[\omega_n^2+\left(\xi-|\bm{g}_{\bm{p}_{\textrm{F}}}|\right)^2\right]^2}-\frac{3}{2\left[\omega_n^2+\left(\xi+|\bm{g}_{\bm{p}_{\textrm{F}}}|\right)^2\right]^2}-\frac{3}{2\left[\omega_n^2+\left(\xi-|\bm{g}_{\bm{p}_{\textrm{F}}}|\right)^2\right]^2}\right.\\\nonumber
&\left.+\frac{2}{\left[\omega_n^2+\left(\xi+|\bm{g}_{\bm{p}_{\textrm{F}}}|\right)^2\right]\left[\omega_n^2+\left(\xi-|\bm{g}_{\bm{p}_{\textrm{F}}}|\right)^2\right]}\right\}\int\frac{d\Omega}{4\pi}\hat{g}_{\bm{p}_{\textrm{F}}i}J_{ij}\left(\bm{p}_{\textrm{F}}\right)\hat{g}_{\bm{p}_{\textrm{F}}k}\\
=&2N\left(E_{\textrm{F}}\right)k_bT\sum_{n=0}^\infty\left[\frac{\pi}{\omega_n^3}-\frac{\pi}{\omega_n\left(\omega_n^2+|\bm{g}_{\bm{p}_{\textrm{F}}}|^2\right)}\right]\left\langle \hat{g}_{\bm{p}_{\textrm{F}}i}J_{ij}\left(\bm{p}_{\textrm{F}}\right)\hat{g}_{\bm{p}_{\textrm{F}}k} \right\rangle,
\end{align}
where we take the infinite-volume limit and the high-density approximation $\frac{1}{V}\sum_{\bm{p}}\rightarrow\int\frac{d^3\bm{p}}{\left(2\pi\right)^3}\rightarrow N\left(E_{\textrm{F}}\right)\int_{-\infty}^{\infty}d\xi\int\frac{d\Omega_{\bm{p}_{\textrm{F}}}}{4\pi^2}$ with $N\left(E_{\textrm{F}}\right)$ the Fermi energy Density of states and $\Omega_{\bm{p}_{\textrm{F}}}$ the solid angle of the Fermi momentum $\bm{p}_{\textrm{F}}$. We set the angle average at the Fermi surface as $\left\langle\right\rangle=\int\frac{d\Omega}{4\pi}$. Substituting Eq. \ref{Green1} and Eq. \ref{Green2} into Eq. \ref{Magnetization} we can get the magnetization $\bm{M}$ as
\begin{align}\nonumber
M_k=&\mu_{\textrm{B}}gN\left(E_{\textrm{F}}\right)\pi k_bT\sum_{q_j}q_j\Delta^\ast_{q_j}\Delta_{q_j}\sum_{n=0}^{\infty}\left[\frac{\left\langle J_{kj}\left(\bm{p}_{\textrm{F}}\right) \right\rangle}{\omega_n\left(\omega_n^2+\bm{g}^2_{\bm{p}_{\textrm{F}}}\right)}+\frac{\left\langle \hat{g}_{\bm{p}_{\textrm{F}}i}J_{ij}\left(\bm{p}_{\textrm{F}}\right)\hat{g}_{\bm{p}_{\textrm{F}}k}\right\rangle\bm{g}^2_{\bm{p}_{\textrm{F}}}}{\omega_n^3\left(\omega_n^2+\bm{g}^2_{\bm{p}_{\textrm{F}}}\right)}\right]\\\nonumber
=&\mu_{\textrm{B}}gN\left(E_{\textrm{F}}\right)\sum_{q_j}q_j\Delta^\ast_{q_j}\Delta_{q_j}\left\{\frac{\left\langle J_{kj}\left(\bm{p}_{\textrm{F}}\right) \right\rangle}{\bm{g}^2_{\bm{p}_{\textrm{F}}}}f\left(\frac{|\bm{g}_{\bm{p}_{\textrm{F}}}|}{\pi k_bT_c}\right)+\left[\frac{7\zeta\left(3\right)}{8\pi^2 k^2_bT^2}-\frac{1}{\bm{g}^2_{\bm{p}_{\textrm{F}}}}f\left(\frac{|\bm{g}_{\bm{p}_{\textrm{F}}}|}{\pi k_bT_c}\right)\right]\left\langle \hat{g}_{\bm{p}_{\textrm{F}}i}J_{ij}\left(\bm{p}_{\textrm{F}}\right)\hat{g}_{\bm{p}_{\textrm{F}}k}\right\rangle\right\},
\end{align}
with $f\left(\rho\right)=\textrm{Re}\sum_{n=0}^{\infty}\left(\frac{1}{2n+1}-\frac{1}{2n+1+i\rho}\right)$ and $\sum_{n=0}^\infty\frac{1}{\left(2n+1\right)^3}=\frac{7\zeta\left(3\right)}{8}$. From the Landau-Ginzburg theory, in the absence of magnetic field we can choose the gauge $\bm{A}=0$ to express the supercurrent density $\bm{J}^{\textrm{S}}$ as
\begin{align}
J^{\textrm{S}}_j=\int d\bm{r} i\hbar\frac{2e}{2m^\ast V}\left[\left(\partial_i\Delta^\ast_{\bm{r}}\right)\Delta_{\bm{r}}-\Delta^\ast_{\bm{r}}\left(\partial_i\Delta_{\bm{r}}\right)\right]=\frac{2e\hbar}{m^\ast}\sum_{q_j}q_j\Delta^\ast_{q_j}\Delta_{q_j},
\end{align}
with $m^\ast$ the effective pairing mass~\cite{Tinkham}. As the supercurrent density $\bm{J}^{\textrm{S}}$ can be normalized by the maximum supercurrent density $J^{\textrm{S}}_{\textrm{max}}=2e\Delta_0^2\frac{2\hbar}{3\sqrt{3}m^\ast\xi}$ with $\xi$ the coherence length and $\Delta_0$ the homogeneous pairing order parameter magnitude~\cite{Tinkham}, we can express $\sum_{q_j}q_j\Delta^\ast_{q_j}\Delta_{q_j}$ as
\begin{align}
\sum_{q_j}q_j\Delta^\ast_{q_j}\Delta_{q_j}=\frac{m^\ast}{2e\hbar}J^{\textrm{S}}_j=\frac{2\Delta_0^2}{3\sqrt{3}\xi}\frac{J^{\textrm{S}}_j}{J^{\textrm{S}}_{\textrm{max}}}.
\end{align}
Eventually, we arrive at the closed form for the magnetization $\bm{M}$
\begin{align}
M_k=T_{kj}\tilde{J}^{\textrm{S}}_j,
\end{align}
with the magnetoelectric pseudo-tensor $T$
\begin{align}\label{T_tensor}
T_{kj}&=\left\{\frac{\Delta_0^2}{\bm{g}^2_{\bm{p}_{\textrm{F}}}}f\left(\frac{|\bm{g}_{\bm{p}_{\textrm{F}}}|}{\pi k_bT_c}\right)\left\langle J_{kj}\left(\bm{p}_{\textrm{F}}\right)\right\rangle+\left[\frac{7\Delta_0^2\zeta\left(3\right)}{8\pi^2k_b^2T_c^2}-\frac{\Delta_0^2}{\bm{g}^2_{\bm{p}_{\textrm{F}}}}f\left(\frac{|\bm{g}_{\bm{p}_{\textrm{F}}}|}{\pi k_bT_c}\right)\right]\left\langle\hat{\bm{g}}_{\bm{p}_{\textrm{F}}k}\hat{\bm{g}}_{\bm{p}_{\textrm{F}}i}J_{ij}\left(\bm{p}_{\textrm{F}}\right)\right\rangle\right\}\frac{2\mu_{\textrm{B}}g}{3\sqrt{3}\xi}N\left(E_{\textrm{F}}\right),
\end{align}
with
\begin{align}
f\left(\rho\right)=\textrm{Re}\sum_{n=0}^\infty\left(\frac{1}{2n+1}-\frac{1}{2n+1+i\rho}\right).
\end{align}
Here $\tilde{\bm{J}}^{\textrm{S}}=\bm{J}^{\textrm{S}}/J^{\textrm{S}}_{\textrm{max}}$ is the normalized supercurrent density.

\section{Total magnetoelectric effect for noncentrosymmetric superconductor with multiple bands}
In the presence of multiple Fermi pockets, the effective Hamiltonian describing the spin-orbit coupled bands away from the Kramers degeneracies is determined by the little group symmetry at the time reversal invariant momenta. Relative to the time reversal invariant momentum $m$, the $l$th pair of spin splitted bands with the singlet pairing can be described by the Bogliubov de Gennes Hamiltonian in the Nambu spinor $\left[c^\dagger_{l, \bm{p}_m+\frac{1}{2}\bm{q}, \uparrow}, c^\dagger_{l, \bm{p}_m+\frac{1}{2}\bm{q}, \downarrow}, c_{l, -\bm{p}_m+\frac{1}{2}\bm{q}, \uparrow}, c_{l, -\bm{p}_m+\frac{1}{2}\bm{q}, \downarrow}\right]^{\textrm{T}}$ as
\begin{align}
H_{m, l}\left(\bm{p}_{m}, \bm{q}\right)=&\begin{pmatrix}
\xi_{l, \bm{p}_m+\frac{1}{2}\bm{q}}+\bm{g}_{l, \bm{p}_m+\frac{1}{2}\bm{q}}\cdot\bm{\sigma} & i\Delta_{l, m, \bm{q}}\sigma_y\\ 
-i\Delta_{l, m, \bm{q}}^\ast\sigma_y & -\xi_{l, -\bm{p}_m+\frac{1}{2}\bm{q}}-\bm{g}_{l, -\bm{p}_m+\frac{1}{2}\bm{q}}\cdot\bm{\sigma}^\ast
\end{pmatrix},
\end{align}
with $\bm{p}_m$ the momentum relative to the time reversal invariant momenta $m$, $\xi_{\bm{p}_m, l}$ the $l$th pair of spin splitted bands dispersion relative to the time reversal invariant momenta $m$, $\bm{g}_{l, \bm{p}_m}$ the spin-orbit coupling pseudovector for the $l$th pair of spin splitted bands relative to the time reversal invariant momenta $m$, and $\Delta_{l, m, \bm{q}}$ the singlet pairing order parameter for the $l$th spin splitted bands relative to the time reversal invariant momenta $m$. The $\bm{q}$ denotes the common net momentum of Cooper pair in the presence of suprecurrent. The form of spin-orbit coupling pseudovector $\bm{g}_{\bm{p}_m, m, l}$ is determined by the little group at the time reversal invariant momenta $m$ as $\bm{g}_{\bm{p}_m, m, l}=\det\left(\hat{R}_m\right)\hat{R}_m\bm{g}_{\hat{R}^{-1}_m\bm{g}_{\bm{p}_m}, m, l}$, with $\hat{R}_{m}$ the little group symmetry operation matrix at the time reversal invariant momenta $m$. Up to the linear order, the form of $\bm{g}_{\bm{p}_m, m, l}$ is present in Table. \ref{magnetoelectric_pseudo-tensor}. Since the effective mean field Hamiltonian has exactly the same form as that of one pair of spin splitted bands, the above derivation for the magnetoelectric susceptibility holds for each pair of Fermi pockets as
\begin{align}\nonumber
T_{kj, m, l}&=\left\{\frac{\Delta_{0, l, m}^2}{\bm{g}^2_{l, \bm{p}_{\textrm{F}, m}}}f\left(\frac{|\bm{g}_{l, \bm{p}_{\textrm{F}, m}}|}{\pi k_bT_c}\right)\left\langle J_{kj, l}\left(\bm{p}_{\textrm{F}, m}\right)\right\rangle\right.\\
&\left.+\left[\frac{7\Delta_{0, l, m}^2\zeta\left(3\right)}{8\pi^2k_b^2T_c^2}-\frac{\Delta_{0, l, m}^2}{\bm{g}^2_{l, \bm{p}_{\textrm{F}, m}}}f\left(\frac{|\bm{g}_{l, \bm{p}_{\textrm{F}, m}}|}{\pi k_bT_c}\right)\right]\left\langle\hat{\bm{g}}_{l, \bm{p}_{\textrm{F}, m}, k}\hat{\bm{g}}_{l, \bm{p}_{\textrm{F}, m}, i}J_{ij, l}\left(\bm{p}_{\textrm{F}, m}\right)\right\rangle\right\}\frac{2\mu_{\textrm{B}}g}{3\sqrt{3}\xi}N_{m, l}\left(E_{\textrm{F}}\right),
\end{align}
with the subscript $m, l$ labelling the corresponding physical quantity at the $l$th pair of Fermi pockets relative to the time reversal invariant momenta $m$. As a result, the total magnetoelectric susceptibility takes into account all the pairing Fermi pockets and reads
\begin{align}
T_{kj}=\sum_{m, l}T_{kj, m, l}.
\end{align}

\section{Details on the symmetry analysis for the magnetoelectric pseudotensor}
The symmetry constraint on the magnetoelectric pseudotensor $T_{ij}$ can be derived from the symmetry transformation of magnetization $\bm{M}$ and the normalized supercurrent density $\tilde{\bm{J}}^{\textrm{S}}$. We know that the magnetoelectric susceptibility $T_{kj}$ connects the magnetization $\bm{M}$ with the normalized supercurrent density $\tilde{\bm{J}}^{\textrm{S}}$ as
\begin{align}
M_k=T_{kj}\tilde{J}^{\textrm{S}}_j.
\end{align}
Under the crystal point group symmetry operation $\hat{R}$, the magnetization $\bm{M}$ transforms as an axial vector $M_k\rightarrow\det\left(\hat{R}\right)\hat{R}_{ki}M_i$, while the normalized supercurrent density $\tilde{\bm{J}}^{\textrm{S}}$ transforms as a polar vector $\tilde{J}_{j}^{\textrm{S}}\rightarrow \hat{R}_{ji}\tilde{J}^{\textrm{S}}_{i}$. As a result, the magnetoelectric susceptibility $T_{kj}$ under the crystal symmetry is subject to the symmetry constraint
\begin{align}\label{Constraint}
T=\det\left(\hat{R}\right)\hat{R}T\hat{R}^{\textrm{T}}.
\end{align}

For a given magnetoelectric pseudotensor $T$, it can be decomposed into the symmetric and anti-symmetric parts $T=\left(T^++T^-\right)$ with $T^{\pm}=\frac{1}{2}\left(T\pm T^{\textrm{T}}\right)$, which transforms independently under symmetry operations. For the anti-symmetric part $T^-$, it can be expressed in terms of a vector $\bm{t}^-$ with $T^-_{ij}=-\epsilon_{ijk}t_k^-$, where $\epsilon_{ijk}$ is the levi-civita tensor. Therefore, in the 10 polar point groups, namely \{C$_n$, C$_{nv}$\} with $n=1, 2, 3, 4, 6$, the presence of polar axis $\bm{c}$ pins the vector $\bm{t}^-$ parallel to $\bm{c}$ and thus the anti-symmetric magnetoelectric tensor $T^-$ keeps invariant under transformation. It contributes to the transverse magnetoelectric response as $\bm{t}^-\times\tilde{\bm{J}}^{\textrm{S}}$ shown in Table \ref{magnetoelectric_pseudo-tensor}. For the symmetric part of the magnetoelectric pseudotensor $T^+$, it survives in the 11 chiral point groups \{O, T, C$_1$, C$_n$, D$_n$\} with $n=2, 3, 4, 6$ and the extra four point groups \{C$_{1v}$, C$_{2v}$, D$_{2d}$, S$_4$\} with improper rotation symmetry. In the chiral point groups, the symmetric $T^+$ represents the magnetization parallel to the supercurrent direction when the supercurrent is applied along the crystal symmetry axis as is shwon in Table \ref{magnetoelectric_pseudo-tensor}. In the point groups \{C$_{1v}$, C$_{2v}$, D$_{2d}$, S$_4$\}, the form of symmetric $T^+$ restricts the longitudinal magnetoelectric response to the plane perpendicular to the principal axis seen from Table \ref{magnetoelectric_pseudo-tensor}.

\begin{table}[ht]
\caption{List of Magnetoelectric pseudotensor $T_{ij}$ in gyrotropic superconductors with spin-orbit coupling $\bm{g}_{\bm{p}}\cdot\bm{\sigma}=\sigma_iv_{ij}p_j$ to the linear order and the candidate gyrotropic superconducting materials} 
\centering 
\begin{tabular}{c c c c c c} 
\hline\hline 
Point group & $v_{ij}$ & $T_{ij}$ & Coordinate notation & Candidate superconductors\\ [0.5ex] 
\hline 
C$_1$ & $\begin{pmatrix}
v_{xx} & v_{xy} & v_{xz} \\ 
v_{yx} & v_{yy} & v_{yz} \\ 
v_{zx} & v_{zy} & v_{zz}
\end{pmatrix}$ & $\begin{pmatrix}
T_{xx} & T_{xy} & T_{xz} \\ 
T_{yx} & T_{yy} & T_{yz} \\ 
T_{zx} & T_{zy} & T_{zz}
\end{pmatrix}$ & arbitrary & - \\
C$_2$ & $\begin{pmatrix}
v_{xx} & v_{xy} & 0 \\ 
v_{yx} & v_{yy} & 0 \\ 
0 & 0 & v_{zz}
\end{pmatrix}$ & $\begin{pmatrix}
T_{xx} & T_{xy} & 0 \\ 
T_{yx} & T_{yy} & 0 \\ 
0 & 0 & T_{zz}
\end{pmatrix}$ & rotation axis along $z$ & UIr, BiPd, ...\\ 
C$_3$ & $\begin{pmatrix}
v_{\parallel} & -v^- & 0 \\ 
v^- & v_{\parallel} & 0 \\ 
0 & 0 & v_{zz}
\end{pmatrix}$ & $\begin{pmatrix}
T_{\parallel} & -T^- & 0 \\ 
T^- & T_{\parallel} & 0 \\ 
0 & 0 & T_{zz}
\end{pmatrix}$ & \tabincell{l}{three fold rotation axis along $z$} & TaRh$_2$B$_2$, NbRh$_2$B$_2$, ... \\
C$_4$ & $\begin{pmatrix}
v_{\parallel} & -v^- & 0 \\ 
v^- & v_{\parallel} & 0 \\ 
0 & 0 & v_{zz}
\end{pmatrix}$ & $\begin{pmatrix}
T_{\parallel} & -T^- & 0 \\ 
T^- & T_{\parallel} & 0 \\ 
0 & 0 & T_{zz}
\end{pmatrix}$ & rotation axis along $z$ & - \\
C$_6$ & $\begin{pmatrix}
v_{\parallel} & -v^- & 0 \\ 
v^- & v_{\parallel} & 0 \\ 
0 & 0 & v_{zz}
\end{pmatrix}$ & $\begin{pmatrix}
T_{\parallel} & -T^- & 0 \\ 
T^- & T_{\parallel} & 0 \\ 
0 & 0 & T_{zz}
\end{pmatrix}$ & rotaton axis long $z$ & - \\
C$_{1v}$ & $\begin{pmatrix}
0 & v_{xy} & 0 \\ 
v_{yx} & 0 & v_{yz} \\ 
0 & v_{zy} & 0
\end{pmatrix}$ & $\begin{pmatrix}
0 & T_{xy} & 0 \\ 
T_{yx} & 0 & T_{yz} \\ 
0 & T_{zy} & 0
\end{pmatrix}$ & in-plane mirror: $y\rightarrow-y$ & \tabincell{l}{Bilayer 1Td-MoTe$_2$, 1Td-WTe$_2$, \\Rh$_2$Ga$_9$, Ir$_2$Ga$_9$, Y$_3$Pt$_4$Ge$_13$, ...} \\
C$_{2v}$ & $\begin{pmatrix}
0 & v_{xy} & 0 \\ 
v_{yx} & 0 & 0 \\ 
0 & 0 & 0
\end{pmatrix}$ & $\begin{pmatrix}
0 & T_{xy} & 0 \\ 
T_{yx} & 0 & 0 \\ 
0 & 0 & 0
\end{pmatrix}$ & \tabincell{c}{rotation axis along $z$,\\ in-plane mirror $x\rightarrow-x, y\rightarrow-y$} & LaNiC$_2$, ThCoC$_2$, ... \\
C$_{3v}$ & $\begin{pmatrix}
0 & -v^- & 0 \\ 
v^- & 0 & 0 \\ 
0 & 0 & 0
\end{pmatrix}$ & $\begin{pmatrix}
0 & -T^- & 0 \\ 
T^- & 0 & 0 \\ 
0 & 0 & 0
\end{pmatrix}$ & \tabincell{l}{rotation axis along $z$, \\in-plane mirror $y\rightarrow-y$} & Li$_2$IrSi$_3$, Gated MoS$_2$, ...\\
C$_{4v}$ & $\begin{pmatrix}
0 & -v^- & 0 \\ 
v^- & 0 & 0 \\ 
0 & 0 & 0
\end{pmatrix}$ & $\begin{pmatrix}
0 & -T^- & 0 \\ 
T^- & 0 & 0 \\ 
0 & 0 & 0
\end{pmatrix}$ & \tabincell{l}{four fold rotation axis along $z$, \\other two $\sigma_v$ mirror $x\rightarrow-x, y\rightarrow-y$} & \tabincell{l}{CePt$_3$Si, LaPt$_3$Si, Ba(Pt, Pd)Si$_3$, \\La(Rh, Pt, Pd, Ir)Si$_3$, Ca(Pt, Ir)Si$_3$, \\Sr(Ni, Pd, Pt)Si$_3$, Sr(Pd, Pt)Ge$_3$, ...} \\
C$_{6v}$ & $\begin{pmatrix}
0 & -T^- & 0 \\ 
T^- & 0 & 0 \\ 
0 & 0 & 0
\end{pmatrix}$ & $\begin{pmatrix}
0 & -T^- & 0 \\ 
T^- & 0 & 0 \\ 
0 & 0 & 0
\end{pmatrix}$ & \tabincell{l}{rotation axis along $z$, \\inplane mirror $y\rightarrow-y$} & Ru$_7$B$_3$, Re$_7$B$_3$, La$_7$Ir$_3$, ... \\
D$_{2d}$ & $\begin{pmatrix}
v_{\parallel} & 0 & 0 \\ 
0 & -v_{\parallel} & 0 \\ 
0 & 0 & 0
\end{pmatrix}$ & $\begin{pmatrix}
T_{\parallel} & 0 & 0 \\ 
0 & -T_{\parallel} & 0 \\ 
0 & 0 & 0
\end{pmatrix}$ & \tabincell{l}{three rotation axis along $x, y, z$, \\in-plane mirror $x\rightarrow y, y\rightarrow x$} & - \\
S$_4$ & $\begin{pmatrix}
v_{\parallel} & v_{\textrm{d}} & 0 \\ 
v_{\textrm{d}} & -v_{\parallel} & 0 \\ 
0 & 0 & 0
\end{pmatrix}$ & $\begin{pmatrix}
T_{\parallel} & T_{\textrm{d}} & 0 \\ 
T_{\textrm{d}} & -T_{\parallel} & 0 \\ 
0 & 0 & 0
\end{pmatrix}$ & improper rotation axis along $z$ & - \\
D$_{2}$ & $\begin{pmatrix}
v_{xx} & 0 & 0 \\ 
0 & v_{yy} & 0 \\ 
0 & 0 & v_{zz}
\end{pmatrix}$ & $\begin{pmatrix}
T_{xx} & 0 & 0 \\ 
0 & T_{yy} & 0 \\ 
0 & 0 & T_{zz}
\end{pmatrix}$ & three rotation axis along $x, y ,z$ & - \\
D$_3$ & $\begin{pmatrix}
v_{\parallel} & 0 & 0 \\ 
0 & v_{\parallel} & 0 \\ 
0 & 0 & v_{zz}
\end{pmatrix}$ & $\begin{pmatrix}
T_{\parallel} & 0 & 0 \\ 
0 & T_{\parallel} & 0 \\ 
0 & 0 & T_{zz}
\end{pmatrix}$ & \tabincell{l}{three fold rotation axis along $z$}  & - \\
D$_4$ & $\begin{pmatrix}
v_{\parallel} & 0 & 0 \\ 
0 & v_{\parallel} & 0 \\ 
0 & 0 & v_{zz}
\end{pmatrix}$  & $\begin{pmatrix}
T_{\parallel} & 0 & 0 \\ 
0 & T_{\parallel} & 0 \\ 
0 & 0 & T_{zz}
\end{pmatrix}$ & \tabincell{l}{four fold rotation axis along $z$} & Zr$_{10}$Sb$_5$Ru,...\\
D$_6$ & $\begin{pmatrix}
v_{\parallel} & 0 & 0 \\ 
0 & v_{\parallel} & 0 \\ 
0 & 0 & v_{zz}
\end{pmatrix}$ & $\begin{pmatrix}
T_{\parallel} & 0 & 0 \\ 
0 & T_{\parallel} & 0 \\ 
0 & 0 & T_{zz}
\end{pmatrix}$ & \tabincell{l}{six fold rotation axis along $z$} & TaSi$_2$, ...\\
T & $\begin{pmatrix}
v_0 & 0 & 0 \\ 
0 & v_0 & 0 \\ 
0 & 0 & v_0
\end{pmatrix}$ & $\begin{pmatrix}
T_0 & 0 & 0 \\ 
0 & T_0 & 0 \\ 
0 & 0 & T_0
\end{pmatrix}$ & $\left(x, y, z\right)$ along crystal axis $\left(a, b ,c\right)$ & AuBe, PdBiSe, NiSbS, ... \\
O & $\begin{pmatrix}
v_0 & 0 & 0 \\ 
0 & v_0 & 0 \\ 
0 & 0 & v_0
\end{pmatrix}$ & $\begin{pmatrix}
T_0 & 0 & 0 \\ 
0 & T_0 & 0 \\ 
0 & 0 & T_0
\end{pmatrix}$ & $\left(x, y, z\right)$ along crystal axis $\left(a, b ,c\right)$ & \tabincell{l}{Li$_2$Pd$_3$B, Li$_2$Pt$_3$B, Mo$_3$Al$_2$C, \\Cr$_2$Re$_3$B, (W, Mo)$_7$Re$_13$(B,C), ...} \\[1ex] 
\hline 
\end{tabular}
\label{magnetoelectric_pseudo-tensor} 
\end{table}

\end{document}